\documentstyle{article}
\begin{document}
\title{
Synthesis of transactinide nuclei in cold fusion reactions using
radioactive beams
}
\author{
Robert Smolanczuk\thanks{Electronic address:  smolan@fuw.edu.pl} 
       \\
	  Theoretical Physics Department,
          So{\l}tan Institute for Nuclear Studies,\\
    Ho\.za 69, PL-00-681 Warszawa, Poland\\ 
        }
\maketitle

\begin{abstract}

Chances of synthesis
of transactinide nuclei in cold fusion reactions (one-neutron-out reactions)
using radioactive beams
are evaluated.
Because intensities of radioactive beams are in most of the cases 
significantly lower than the ones of the stable beams, reactions with 
the highest radioactive beam intensities for the particular elements
are considered. The results are
compared with the recent ones obtained by Loveland \cite{lov07}
who investigated the same nuclei.

\end{abstract}

Since cold fusion reactions based on $^{208}$Pb and $^{209}$Bi targets
have been proposed \cite{og75b}
elements from Bh ($Z=107$) up to 112 have been discovered
by using these reactions \cite{hofmue}. 
Recently also synthesis of element 113 has
been reported using this method \cite{mor04b}.
All these nuclei have been obtained using stable projectiles.
The objective of the present paper is to evaluate chances of production, 
in other
words production rates, of more neutron reach nuclei which may be
obtained in cold fusion reactions using neutron-rich radioactive-ion beams.
For this purpose we use beam intensities predicted on the
Rare Isotope Accelerator
(RIA) webpage, which are listed in Ref.\cite{lov07},
and formation cross sections calculated
in our model introduced in Ref.\cite{smo08}, where the formation cross
sections for cold fusion reactions using stable beams have been reproduced. 

The formation
cross section for optimal bombarding energy corresponding to the maximum of 
the excitation function we calculate by using the model formula
\begin{equation}
\sigma_{1n}=
c_1 \sigma_{\rm class} P_{\rm CN} (\frac{\Gamma_n}{\Gamma_f})_{l=0}=
\sigma_{capt} P_{CN} (\frac{\Gamma_n}{\Gamma_f})_{l=0}=
\frac{1}{2} \pi R_B^2 (1-\frac{V_B}{E_{cm}}) \cdot
  \exp{\{}-a [R_B-R_{inner} ]{\}} 
\cdot
\exp (\frac{B_f+\Delta_{sd}}{T_{sd}}-
      \frac{S_n+\Delta_{eq}}{T_{eq}}).
\label{simple}
\end{equation}
Formula (\ref{simple}) consists of four factors. 
The first factor is a coefficient $c_1$ which originates from integration
over partial waves and for cold fusion
reactions leading to transactinides has been evaluated in Ref.\cite{smo08}
to be equal to approximatly 
0.5. The following three factors are
the classical cross section $\sigma_{\rm class}$,
the formation probability of the compound nucleus (the so-called fusion
hindrence) $P_{CN}$ and the survival probability of the compound nucleus for
the maximum of the excitation function and for zero angular momentum
$(\frac{\Gamma_n}{\Gamma_f})_{l=0}$. The so-called capture cross section
is equal to $c_1 \cdot \sigma{class}$.

The clasical cross section (the
considered reactions are overbarrier reactions, see Table I)
depends on
radial coordinate of the average fusion barrier $R_B$, the height of the 
average fusion
barrier $V_B$ and the optimal
bombarding energy in the center-of-mass system $E_{cm}$
corresponding to the maximum of the excitation function.
The fusion barrier and its position are calculated by using the folding
potential with built-in dependence on the thickness of nuclear surface
and the separation energy of the least bound nucleon. The separation energy
is calculated using masses from Ref.\cite{wap03}. Full details of
the potential are given in Ref.\cite{smo08}.

The formation probability of the compound nucleus $P_{CN}$ 
depends on the difference
between radial coordinates of the average fusion barrier $R_B$ and the inner
barrier $R_{inner}$.
The larger distance between both barriers the less likely
compound nucleus formation because of the competing quasifission
process.
$P_{CN}$ is independent of energy in the energy range of interest
what one can deduce from the experimental data obtained for
$^{220}$Th \cite{hin02}. We also assumed that $P_{CN}$ is independent
of angular momentum.
The position of the inner barrier $R_{inner}=b_{inner}/x_f$
is assumed to be inversly proportional to the fissility parameter
$x_f=4Z_T Z_P/[A_T^{1/3} A_P^{1/3} (A_T^{1/3}+A_P^{1/3})]$,
where $Z_T$, $A_T$ and $Z_P$, $A_P$ are the atomic and mass numbers
of the target $T$ and the projectile $P$, respectively.
The constant $b_{inner}=427.69$ fm was determined assuming that
the reaction $^{208}$Pb($^{48}$Ca,$1n$)$^{255}$No is unhindered 
($P_{CN}=1$) because of very low quasifission rates measured for
this reaction \cite{itkis}. The constant $a=5.03$ fm$^{-1}$ was adjusted
to the maxima of the
following measured
exitation functions for one-neutron-out reactions
$^{208}$Pb($^{50}$Ti,$1n$)$^{257}$Rf, 
$^{209}$Bi($^{50}$Ti,$1n$)$^{258}$Db, 
$^{208}$Pb($^{54}$Cr,$1n$)$^{261}$Sg, 
$^{208}$Pb($^{58}$Fe,$1n$)$^{265}$Hs and 
$^{208}$Pb($^{64}$Ni,$1n$)$^{271}$Ds \cite{hof04}.

The survival probability at the maximum of the excitation function
$(\Gamma_n /\Gamma_f)_{l=0}$ depends on effective tresholds for fission
$B_f+\Delta_{sd}$, neutron emission $S_n+\Delta_{eq}$ and temperatures 
at the saddle point of the compound nucleus $T_{sd}$ and at the 
equilibrium point of the evaporation residue $T_{eq}$.
Here, $B_f$ is the static fission barrier, $S_n$ is the
neutron separation energy,
$\Delta_{sd}$ and $\Delta_{eq}$ are the energy shifts
in the saddle-point of the compound nucleus
and the equilibrium configuration of $1n$-evaporation residue,
respectively. They are used to take into account differences between level
densities for even-even, odd and odd-odd nuclei.
The energy shift $\Delta_{sd}$ is equal to 
$24/\sqrt{A}$, $12/\sqrt{A}$ and 0 for
the even-even, odd and odd-odd compound nucleus and
$\Delta_{eq}=24/\sqrt{A-1}$, $12/\sqrt{A-1}$ and 0 for
the even-even, odd and odd-odd $1n$-evaporation residue, respectively.
$A$ is the mass number of the compound nucleus.

Influence of shell effects on the survival probability
was taken implicitly into account through different temperatures
in the transitional states for fission and neutron emission.
Temperature at the equilibrium point of the evaporation residue
$T_{eq}=(\frac{d}{dE^*}ln\rho)^{-1}$ is smaller than at the
saddle point of the compound nucleus
$T_{sd}$ because the level density $\rho(E^*)$ in the equilibrium
configuration increases faster 
with increasing excitation energy $E^*$ in comparison with the level 
density in the saddle-point configuration of the compound nucleus.
This is due to thermal damping of the strong 
ground-state shell-effect. In the saddle point, there is no shell effect 
or it is much weaker than in the equilibrium configuration. 
In the calculation $T_{sd}=1$ MeV, $T_{eq}=0.55$ MeV for the even-even
and odd-N compound nucleus and $T_{eq}=0.5$ MeV for odd-Z and odd-odd
compound nucleus were used. In Ref.\cite{smo08} it was checked out that
the values of survival probabilities obtained with these temperatures
are close to those obtained 
with shell damping of the experimental level density parameter
\cite{mug98} described by the Ignatyuk, Smirenkin and Tishin formula
\cite{ign75}.

The optimal bombarding energy in the center-of-mass system
$E_{cm}$ corresponding to the maximum of
the excitation function is calculated as the sum of the absolute value of 
$Q$-value necessary to form an unexcited compound nucleus, the effective
threshold for one neutron emission and the effective threshold for
next decay process, which for the evaporation residue is fission
because for transactinides fission barriers are smaller than neutron
separation energies.

The $Q$-value was calculated by subtracting from 
the measured masses \cite{wap03} of the reaction partners
the mass of the compound nucleus that was calculated
in the macroscopic-microscopic model \cite{phd,smo97}.
The static fission barrier height and the 
neutron separation energy are obtained in the macroscopic-microscopic
model \cite{sss94,phd,smo97}. 

Obtained results in the present paper and in the paper by Loveland
who investigated the same nuclei
\cite{lov07}, as well as beam intensities
given in Ref.\cite{lov07} after the 
RIA webpage are collected in Table I.
Because intensities of radioactive beams are in most of the cases 
significantly lower than the ones of the stable beams, reactions using 
the highest radioactive beam intensities for the particular elements
are considered. 
One can see that our cross section of $^{208}$Pb($^{51}$Ti,$1n$)$^{258}$Rf
seems to be to large. $^{258}$Rf has only one neutron more than the
nucleus obtained by using stable $^{50}$Ti beam. From systematics we know
that this should lead
to increase of the cross section by a factor smaller than one
order of magnitude
whereis the calculation gives the increse of the cross section by two
orders of magnitude. Possible reason for this is both 
overesimation of the fission
barrier and underestimation of the neutron separation energy in the
macroscopic-microscopic model for the compound nucleus $^{259}$Rf.

As stated in Ref.\cite{lov07} the cross section times beam intensity factor
governs the choice of reactions. 
To have an idea how the cross section times beam intensity factor
works we use as the reference the reaction 
$^{208}$Pb($^{70}$Zn,$1n$)$^{277}112$
with the stable beam $^{70}$Zn with the intensity of $3 \cdot 10^{12}$ 
for which the cross section of 0.5~pb was measured \cite{hof04}.
These numbers correspond to the production rate of 23 days/atom.

The best reaction taking into account the cross section times beam intensity
factor and disregarding overestmated result for
$^{208}$Pb($^{51}$Ti,$1n$)$^{258}$Rf is 
$^{208}$Pb($^{54}$V,$1n$)$^{261}$Db, the same reaction as indicated
in Ref.\cite{lov07}. We obtain the production rate 0.13 days/atom.

Our obtained results are generaly smaller than those obtained
in Ref.\cite{lov07} except for 
$^{208}$Pb($^{54}$V,$1n$)$^{261}$Db.
Synthesis of lower $Z$ (106-107, 109-111) 
transactinides using neutron-rich radiactive
ion beams is most likely if beam intensities 
given in Ref.\cite{lov07} after the RIA webpage are predicted 
correctly. 
Production rates are of the order of days/atom for these elements
except for element Rg (Z=111) for which production rate of almost 23 days/atom
have been obtained.

The synthesis of higher $Z$ transactinide nuclei 
by means of the neutron-rich radioactive-ion beams
seems to be practically excluded using present-day 
technology because of small cross sections and low beam intensities 
(small cross section times beam intensity factor leading to
low production rates).
Our results are even more pesymistic in this regard in comparison
with the results obtained by Loveland \cite{lov07}.
We obtain production rates even up to two orders of magnitude
smaller than that observed for the reference reaction and for the
case of
$^{208}$Pb($^{78}$As,$1n$)$^{285}$115 even up to three orders of magnitude.
The only exeption is the reaction 
$^{208}$Pb($^{91}$Kr,$1n$)$^{298}118$. In this case the predicted
beam intensity is high enough to produce the nuclei at the
production rate of 23 atoms/day to obtain the isotope of element 118
with the neutron number 180 very close to the neutron magic number 184.

\newpage
TABLE~I. The optimal bombarding energy in the center-of-mass-system
$E_{cm}$, the average fusion barrier height 
$V_B$, the radial coordinates of the average
fusion barrier $R_B$ and the inner barrier $R_{inner}$, the
capture cross section $\sigma_{capt}$, the formation probability of
the compound nucleus $P_{CN}$, the neutron-to-fission-width-ratio
for zero angular momentum $(\Gamma_n/\Gamma_f)_{l=0}$ and the
formation cross sections for $1n$-evaporation residue 
$\sigma_{1n}$ calculated
in the present paper and in the paper by Loveland $\sigma_{1n}^L$
\cite{lov07}
as well as the beam intensity given in Ref.\cite{lov07} after the RIA
webpage.
\begin{center}
\begin{tabular}{c|c|c|c|c|c|c|c|l|l|c}
\hline
\hline
Reaction &  
   $E_{cm}$  &   $V_B$  &  $R_B$ &  $R_{inner}$  &
   $\sigma_{capt}$   &  $P_{CN}$            & $(\Gamma_n/\Gamma_f)_{l=0}$ &
   $\sigma_{1n}$     & $\sigma_{1n}^L$    &  $\Psi$    \\
         &  
     MeV     &   MeV    &   fm   &   fm          &
     mb              &                      &                             &
     pb              &   pb              &   ions/s                       \\
\hline
$^{208}$Pb($^{51}$Ti,$1n$)$^{258}$Rf    &
     185.8   &   182.2  &  13.57 &  12.55        &  
      56             & $5.8\cdot 10^{-3}$   &    $3.6\cdot 10^{-3}$       &    
     1200000     &   5190             &   $4.0\cdot 10^9$               \\
$^{208}$Pb($^{54}$V,$1n$)$^{261}$Db    &
     192.7   &   189.4  &  13.64 &  12.32        &  
      49             & $1.3\cdot 10^{-3}$   &    $1.2\cdot 10^{-3}$       &    
     74000    &      11700            &   $3.6\cdot 10^9$               \\
$^{208}$Pb($^{57}$Cr,$1n$)$^{264}$Sg   &
     201.9   &   196.3  &  13.73 &  12.11        &  
      82             & $2.9\cdot 10^{-4}$   &    $1.8\cdot 10^{-3}$       &    
     4300    &       8890             & $2.8\cdot 10^9$                      \\
$^{208}$Pb($^{58}$Mn,$1n$)$^{265}$Bh   &
     210.3   &   204.8  &  13.71 &  11.72        &  
      78             & $4.5\cdot 10^{-5}$   &    $5.0\cdot 10^{-4}$       &    
     1700    &       3210             &   $8.0\cdot 10^9$                \\
$^{208}$Pb($^{61}$Fe,$1n$)$^{268}$Hs   &
     220.2   &   211.6  &  13.80 &  11.54        &  
     117             & $1.1\cdot 10^{-5}$   &    $9.4\cdot 10^{-4}$       &    
     120    &        3380               &   $4.0\cdot 10^9$            \\
$^{208}$Pb($^{58}$Co,$1n$)$^{265}$Mt   &
     224.9   &   222.1  &  13.66 &  10.85        &  
      38             & $7.3\cdot 10^{-7}$   &    $5.9\cdot 10^{-5}$       &    
     1.6   &         2.4                &   $4.0\cdot 10^{12}$            \\
$^{208}$Pb($^{65}$Ni,$1n$)$^{272}$Ds   &
     239.1   &   226.9  &  13.85 &  11.03        &  
     154             & $7.0\cdot 10^{-7}$   &    $9.6\cdot 10^{-4}$       &    
     100   &         497               &   $3.6\cdot 10^{10}$                \\
$^{208}$Pb($^{66}$Cu,$1n$)$^{273}$Rg   &
     246.6   &   235.2  &  13.85 &  10.73        &  
     139             & $1.5\cdot 10^{-7}$   &    $2.6\cdot 10^{-4}$       &    
     5.5   &         60                &   $2.8\cdot 10^{11}$              \\
$^{209}$Bi($^{66}$Cu,$1n$)$^{274}$112   &
     248.8   &   237.9  &  13.85 &  10.63        &  
     132             & $9.1\cdot 10^{-8}$   &    $1.3\cdot 10^{-4}$       &    
     1.5  &          5.7                &  $2.8\cdot 10^{11}$               \\
$^{208}$Pb($^{74}$Ga,$1n$)$^{281}$113   &
     265.1   &   248.5  &  14.00 &  10.59        &  
     192             & $3.6\cdot 10^{-8}$   &    $5.9\cdot 10^{-5}$       &    
     0.41   &        18.9               &   $1.2\cdot 10^{11}$           \\
$^{209}$Bi($^{74}$Ga,$1n$)$^{282}$114   &
     267.4   &   251.4  &  14.01 &  10.49        &  
     184             & $2.1\cdot 10^{-8}$   &    $6.1\cdot 10^{-5}$       &    
     0.23  &         34.5               &   $1.2\cdot 10^{11}$       \\
$^{208}$Pb($^{78}$As,$1n$)$^{285}$115   &
     283.7   &   263.6  &  14.06 &  10.20        &  
     220             & $3.7\cdot 10^{-9}$   &    $1.7\cdot 10^{-5}$       &    
     0.014  &        7.6               &   $8.0\cdot 10^{10}$        \\
$^{208}$Pb($^{85}$Se,$1n$)$^{292}$116   &
     297.8   &   267.8  &  14.25 &  10.31        &  
     322             & $2.5\cdot 10^{-9}$  &    $1.4\cdot 10^{-3}$       &    
     1.1 &           17.6              &   $1.6\cdot 10^{10}$              \\
$^{208}$Pb($^{89}$Br,$1n$)$^{296}$117   &
     303.7   &   275.6  &  14.26 &  10.24        &  
     296             & $1.6\cdot 10^{-9}$  &    $1.1\cdot 10^{-3}$       &    
     0.55 &          3.4              &   $8.0\cdot 10^{10}$               \\
$^{208}$Pb($^{91}$Kr,$1n$)$^{298}$118   &
     314.4   &   281.6  &  14.34 &  10.06        &  
     337             & $4.5\cdot 10^{-10}$  &    $3.6\cdot 10^{-3}$       &    
     0.55 &          0.7               &   $2.8\cdot 10^{12}$               \\
$^{208}$Pb($^{90}$Rb,$1n$)$^{297}$119   &
     326.1   &   290.7  &  14.28 &   9.74        &  
     347             & $1.2\cdot 10^{-10}$  &    $5.4\cdot 10^{-4}$       &    
     0.022 &         0.1            &   $3.6\cdot 10^{12}$                  \\
\hline
\hline
\end{tabular}
\end{center}

\end{document}